\documentclass[12pt]{article}
\usepackage[dvips]{epsfig}

\begin{document}

\title{Confinement of neutral fermions by a pseudoscalar double-step potential in
(1+1) dimensions}
\date{}
\author{Antonio S. de Castro\footnote{castro@feg.unesp.br} and Wiliam G. Pereira \\
\\
UNESP - Campus de Guaratinguet\'{a}\\
Departamento de F\'{\i}sica e Qu\'{\i}mica\\
Caixa Postal 205\\
12516-410 Guaratinguet\'{a} SP - Brasil}
\maketitle

\begin{abstract}
The problem of confinement of neutral fermions in two-dimensional space-time
is approached with a pseudoscalar double-step potential in the Dirac
equation. Bound-state solutions are obtained when the coupling is of
sufficient intensity. The confinement is made plausible by arguments based
on effective mass and anomalous magnetic interaction.
\end{abstract}

\pagebreak

In a recent Letter the problem of confinement of fermions by mixed
vector-scalar linear potentials in 1+1 dimensions with the Dirac equation
was approached \cite{asc}. There it was found that relativistic confining
potentials may result in no bound-state solutions in the nonrelativistic
limit. In the present Letter some aspects of the spectrum of a fermion in a
double-step potential with pseudoscalar coupling in a 1+1 dimensional
space-time are analyzed. This essentially relativistic problem gives rise to
confinement furnishing another opportunity to come up with that there exist
relativistic confining potentials providing no bound-state solutions in the
nonrelativistic limit. To our knowledge, this sort of problem has not been
solved before.

Let us begin by presenting the Dirac equation in 1+1 dimensions. In the
presence of a time-independent potential the 1+1 dimensional
time-independent Dirac equation for a fermion of rest mass $m$ reads
\begin{equation}
\mathcal{H}\Psi =E\Psi  \label{eq1}
\end{equation}

\begin{equation}
\mathcal{H}=c\alpha p+\beta mc^{2}+\mathcal{V}  \label{eq1a}
\end{equation}

\noindent where $E$ is the energy of the fermion, $c$ is the velocity of
light and $p$ is the momentum operator. $\alpha $ and $\beta $ are Hermitian
square matrices satisfying the relations $\alpha ^{2}=\beta ^{2}=1$, $%
\left\{ \alpha ,\beta \right\} =0$. From the last two relations it steams
that both $\alpha $ and $\beta $ are traceless and have eigenvalues equal to
$-$1, so that one can conclude that $\alpha $ and $\beta $ are
even-dimensional matrices. One can choose the 2$\times $2 Pauli matrices
satisfying the same algebra as $\alpha $ and $\beta $, resulting in a
2-component spinor $\Psi $. The positive definite function $|\Psi |^{2}=\Psi
^{\dagger }\Psi $, satisfying a continuity equation, is interpreted as a
probability position density and its norm is a constant of motion. This
interpretation is completely satisfactory for single-particle states \cite
{tha}. We use $\alpha =\sigma _{1}$ and $\beta =\sigma _{3}$. For the
potential matrix we consider
\begin{equation}
\mathcal{V}=1V_{t}+\beta V_{s}+\alpha V_{e}+\beta \gamma ^{5}V_{p}
\label{eq2}
\end{equation}

\noindent where $1$ stands for the 2$\times $2 identity matrix and $\beta
\gamma ^{5}=\sigma _{2}$. This is the most general combination of Lorentz
structures for the potential matrix because there are only four linearly
independent 2$\times $2 matrices. The subscripts for the terms of potential
denote their properties under a Lorentz transformation: $t$ and $e$ for the
time and space components of the 2-vector potential, $s$ and $p$ for the
scalar and pseudoscalar terms, respectively. It is worth to note that the
Dirac equation is invariant under $x\rightarrow -x$ if $V_{e}(x)$ and $%
V_{p}(x)$ change sign whereas $V_{t}(x)$ and $V_{s}(x)$ remain the same.
This is because the parity operator $P=\exp (i\theta )P_{0}\sigma _{3}$,
where $\theta $ is a constant phase and $P_{0}$ changes $x$ into $-x$,
changes sign of $\alpha $ and $\beta \gamma ^{5}$ but not of $1$ and $\beta $%
.

Defining the spinor $\psi $ as
\begin{equation}
\psi =\exp \left( \frac{i}{\hbar }\Lambda \right) \Psi  \label{eq5}
\end{equation}

\noindent where
\begin{equation}
\Lambda (x)=\int^{x}dx^{\prime }\frac{V_{e}(x^{\prime })}{c}  \label{eq6}
\end{equation}

\noindent the space component of the vector potential is gauged away

\begin{equation}
\left( p+\frac{V_{e}}{c}\right) \Psi =\exp \left( \frac{i}{\hbar }\Lambda
\right) p\psi  \label{eq7}
\end{equation}

\noindent so that the time-independent Dirac equation can be rewritten as
follows:

\begin{equation}
H\psi =E\psi  \label{eq7a}
\end{equation}

\begin{equation}
H=\sigma _{1}cp+\sigma _{2}V_{p}+\sigma _{3}\left( mc^{2}+V_{s}\right)
+1V_{t}  \label{eq8}
\end{equation}

\noindent showing that the space component of a vector potential only
contributes to change the spinors by a local phase factor.

Provided that the spinor is written in terms of the upper and the lower
components
\begin{equation}
\psi =\left(
\begin{array}{c}
\phi \\
\chi
\end{array}
\right)  \label{eq8a}
\end{equation}

\noindent the Dirac equation decomposes into :

\begin{eqnarray}
\left( V_{t}-E+V_{s}+mc^{2}\right) \phi (x) &=&i\hbar c\chi ^{\prime
}(x)+iV_{p}\chi (x)  \nonumber \\
&&  \label{eq8b} \\
\left( V_{t}-E-V_{s}-mc^{2}\right) \chi (x) &=&i\hbar c\phi ^{\prime
}(x)-iV_{p}\phi (x)  \nonumber
\end{eqnarray}

\noindent where the prime denotes differentiation with respect to $x$. In
terms of $\phi $ and $\chi $ the spinor is normalized as $\int_{-\infty
}^{+\infty }dx\left( |\phi |^{2}+|\chi |^{2}\right) =1$, so that $\phi $ and
$\chi $ are square integrable functions. It is clear from the pair of
coupled first-order differential equations (\ref{eq8b}) that both $\phi (x)$
and $\chi (x)$ have opposite parities and must be discontinuous wherever the
potential undergoes an infinite jump. In the nonrelativistic approximation
(potential energies small compared to the rest mass) Eq. (\ref{eq8b}) becomes

\begin{equation}
\chi =\frac{p}{2mc}\phi  \label{eq8c}
\end{equation}

\begin{equation}
\left( -\frac{\hbar ^{2}}{2m}\frac{d^{2}}{dx^{2}}+V_{t}+V_{s}\right) \phi
=\left( E-mc^{2}\right) \phi  \label{eq8d}
\end{equation}

\noindent Eq. (\ref{eq8c}) shows that $\chi $ if of order $v/c<<1$ relative
to $\phi $ and Eq. (\ref{eq8d}) shows that $\phi $ obeys the Schr\"{o}dinger
equation without any contribution from the pseudoscalar potential.

Now, let us choose the pseudoscalar double-step potential

\begin{equation}
V_{p}=\left\{
\begin{array}{cc}
-V_{0}, & x<-a/2 \\
0, & |x|<a/2 \\
+V_{0}, & x>a/2
\end{array}
\right.  \label{eq40}
\end{equation}

\noindent with $V_{0}$ and $a$ defined to be positive real numbers. Then the
general solutions of the Dirac equation can be written as

\begin{equation}
\psi (x)=\left\{
\begin{array}{cc}
A\left(
\begin{array}{c}
1 \\
-i\mu _{1}^{\left( +\right) }
\end{array}
\right) e^{+ik_{1}x}+A^{^{\prime }}\left(
\begin{array}{c}
1 \\
-i\mu _{1}^{\left( -\right) }
\end{array}
\right) e^{-ik_{1}x}, & x<-a/2 \\
&  \\
B\left(
\begin{array}{c}
1 \\
\mu _{2}
\end{array}
\right) e^{+ik_{2}x}+B^{^{\prime }}\left(
\begin{array}{c}
1 \\
-\mu _{2}
\end{array}
\right) e^{-ik_{2}x}, & |x|<a/2 \\
&  \\
C\left(
\begin{array}{c}
1 \\
i\mu _{1}^{\left( -\right) }
\end{array}
\right) e^{+ik_{1}x}+C^{^{\prime }}\left(
\begin{array}{c}
1 \\
i\mu _{1}^{\left( +\right) }
\end{array}
\right) e^{-ik_{1}x}, & x>a/2
\end{array}
\right.  \label{eq50}
\end{equation}

\noindent where

\begin{eqnarray}
\left( \hbar ck_{1}\right) ^{2} &=&E^{2}-m^{2}c^{4}-V_{0}^{2}  \nonumber \\
&&  \nonumber \\
\left( \hbar ck_{2}\right) ^{2} &=&E^{2}-m^{2}c^{4}  \nonumber \\
&&  \label{eq60} \\
\mu _{1}^{\left( \pm \right) } &=&\frac{\pm i\hbar ck_{1}+V_{0}}{E+mc^{2}}
\nonumber \\
&&  \nonumber \\
\mu _{2} &=&\frac{\hbar ck_{2}}{E+mc^{2}}  \nonumber
\end{eqnarray}

\noindent

\noindent

\noindent If $E^{2}<m^{2}c^{4}+V_{0}^{2}$ then $k_{1}$ is an imaginary
number, hence the spinors in the region $|x|>a/2$ will describe attenuated
exponentially wavefunctions, the necessary condition for the existence of
bound states. Otherwise, the spinors will describe scattering states.

We turn our attention to the bound states solutions. When the energy is such
that $m^{2}c^{4}<E^{2}<m^{2}c^{4}+V_{0}^{2}$ one must have $A=C^{^{\prime
}}=0$ in order to obtain normalizable wavefunctions in the region $|x|>a/2$.
From the requirement that the spinor must be continuous at $x=\pm a/2$ one
obtains four relations involving the four remaining coefficients of the
spinor, however one of them is to be determined by the normalization. The
joining condition of $\psi (x)$ at $x=\pm a/2$ leads to the quantization
conditions

\begin{equation}
\begin{array}{cc}
&  \\
\tan \left( k_{2}a/2\right) =\mu _{1}^{\left( -\right) }/\mu _{2}, & \phi \
\textrm{even}\;\;\;(B=B^{^{\prime }},\;C=+A^{^{\prime }}) \\
&  \\
-\cot \left( k_{2}a/2\right) =\mu _{1}^{\left( -\right) }/\mu _{2}, & \phi \
\textrm{odd}\;\;\;(B=-B^{^{\prime }},\;C=-A^{^{\prime }})
\end{array}
\label{eq70}
\end{equation}

\begin{figure}[tbp]
\begin{center}
\begin{minipage}{20\linewidth}
\epsfig{file=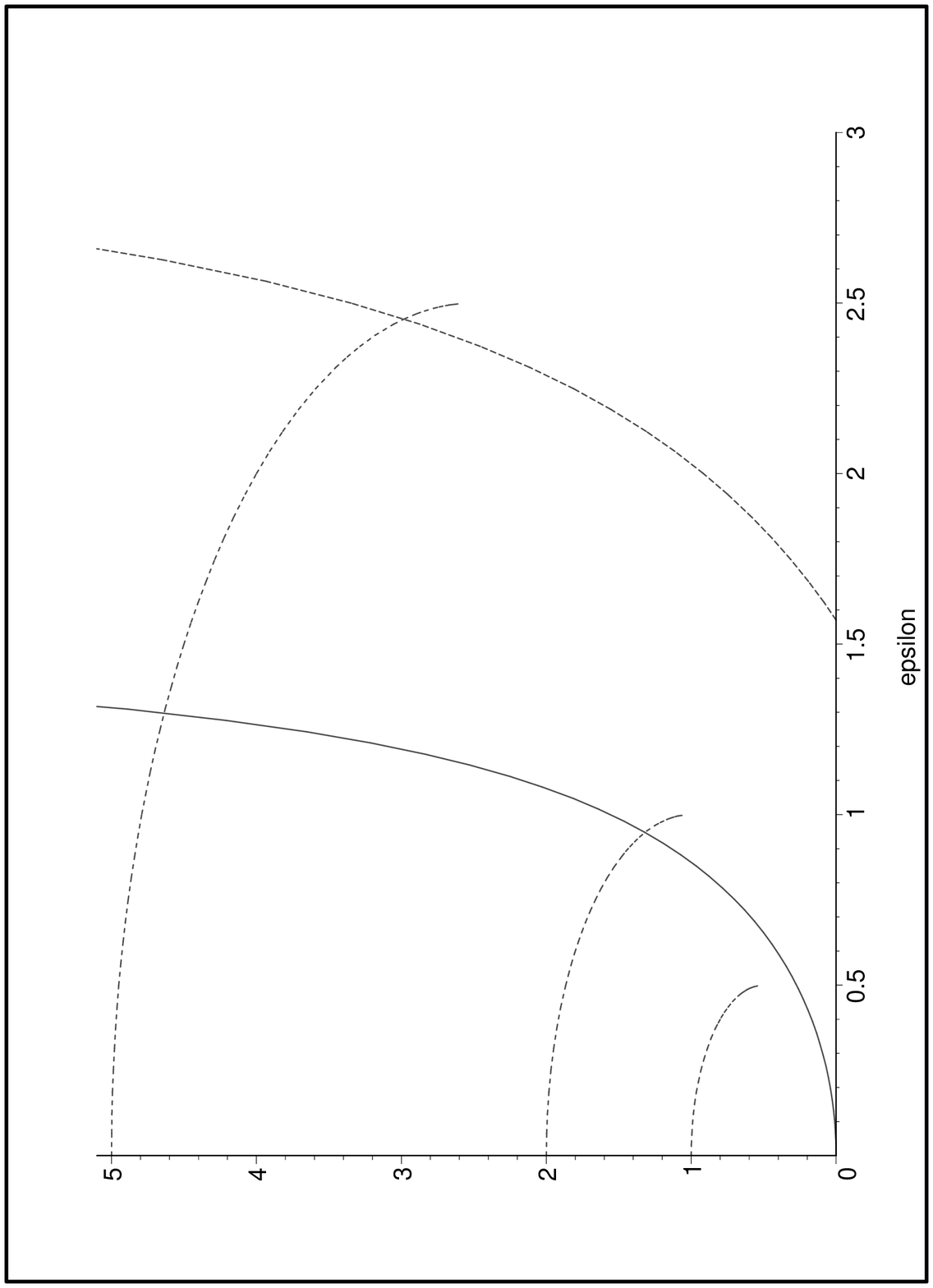}
\end{minipage}
\end{center}
\caption{Graphical solution of Eq. (\ref{eq100}) for three values of $%
\vartheta $. The solid line represents $\varepsilon \tan \left( \varepsilon \right) $
and the dashed line represents $-\varepsilon \cot \left( \varepsilon \right) $. The
dashed-dotted lines represent the right-handed side of Eq. (\ref{eq100}).}
\label{fig:F1}
\end{figure}

\noindent It should be noted from (\ref{eq70}) that the eigenvalues of the
Dirac equation are symmetrical about $E=0$. Making the change of variables

\begin{eqnarray}
\varepsilon &=&\frac{k_{2}a}{2}  \nonumber \\
&&  \label{eq90} \\
\vartheta &=&\frac{V_{0}a}{2\hbar c}  \nonumber
\end{eqnarray}

\noindent the quantization conditions in terms of these dimensionless
quantities take the form

\begin{equation}
\begin{array}{cc}
\varepsilon \tan \left( \varepsilon \right) =\sqrt{\vartheta
^{2}-\varepsilon ^{2}}+\vartheta , & \phi \ \textrm{even} \\
&  \\
-\varepsilon \cot \left( \varepsilon \right) =\sqrt{\vartheta ^{2}-\varepsilon
^{2}}+\vartheta , & \phi \ \textrm{odd}
\end{array}
\label{eq100}
\end{equation}

\noindent \noindent \noindent with the restriction $0<\varepsilon <\vartheta $. The
solutions of these transcendental equations are to be found from numerical or
graphical methods. The graphical method is illustrated in Figure \ref{fig:F1} for
three values of $\upsilon $.The solutions for $\phi $ even
(odd) are given by the intersection of the curve represented by $\sqrt{%
\vartheta ^{2}-\varepsilon ^{2}}+\vartheta $ with the curve represented by $%
\varepsilon \tan \left( \varepsilon \right) $ ($-\varepsilon \cot \left(
\varepsilon \right) $). We can see immediately that it needs critical
values, corresponding to $\tan \left( \varepsilon \right) =\pm 1$, for
obtaining bound states. If $\vartheta $ is larger than the critical values
there will be a finite sequence of bound states with alternating parities.
The ground-state solution will correspond to $\phi $ even. When $\upsilon $
approaches infinity the intersections will occur at the asymptotes of $%
\varepsilon \tan \left( \varepsilon \right) $ and $-\varepsilon \cot \left(
\varepsilon \right) $ so that the eigenvalues will be given by

\begin{equation}
\varepsilon _{n}^{\left( \infty \right) }=\frac{n\pi }{2},\qquad
n=1,2,3,\ldots  \label{eq110}
\end{equation}

\noindent Then, if

\begin{equation}
\frac{\pi }{4}\leq \vartheta <\left( 2N+1\right) \frac{\pi }{4}
\label{eq110a}
\end{equation}

\noindent there will be $N$ bound states with eigenvalues given by

\begin{equation}
\left( 2N-1\right) \frac{\pi }{4}\leq \varepsilon _{N}<N\frac{\pi }{2}
\label{eq110b}
\end{equation}

As the potential is a double step, one should not expect the existence of
bound states, and it follows that such bound states are consequence of the
peculiar coupling in the Dirac equation. The existence of bound states can
be easily understood if one realizes that the wave number $k_{1}$ at the top
line of (\ref{eq60}) can be obtained from the free wave number $k_{2}$ by
the substitution $m\rightarrow \sqrt{m^{2}+V_{0}^{2}/c^{4}}$, a result
independent of the sign of $V_{0}$. This fact shows that this kind of
potential couples to the mass of the fermion and consequently it couples to
the positive-energy component of the spinor in the same way it couples to
the negative-energy component. The energy spectrum for a free fermion ranges
continuously from $-mc^{2}$ to $-\infty $ as well as from $mc^{2}$ to $%
\infty $ and the double-step potential contributes to enhance the energy gap
bringing into being at $x=\pm a/2$ an ascending step for the positive-energy
solutions and a descending step for the negative-energy solutions. In this
way there is no room for transitions from positive-energy solutions to
negative-energy ones. It means that Klein\'{}s paradox does not come to the
scenario.

The Dirac equation with a nonvector potential, or a vector potential
contaminated with some scalar or pseudoscalar coupling, is not invariant
under $V\rightarrow V+const.$, this is so because only the vector potential
couples to the positive-energies in the same way it couples to the
negative-ones, whereas nonvector contaminants couple to the mass of the
fermion. Therefore, if there is any nonvector coupling the absolute values
of the energy will have physical significance and the freedom to choose a
zero-energy will be lost.

The problem approached in this Letter is just a simple example of a
pseudoscalar potential which is able to trap a neutral fermion, but it turns
out that other one-dimensional potentials can do the same. The linear
potential known as the Dirac oscillator is one of such examples.

Moreno and Zentella \cite{mz} showed that the term $-im\omega \beta \vec{r}$
in the Dirac equation corresponds to an anomalous magnetic interaction.
Ben\'{i}tez \textit{et al. } \cite{be} also showed that the Dirac oscillator
may be considered as describing a neutral fermion with anomalous magnetic
moment interacting with a static and spherically symmetric linear electric
field. Later, Lin \cite{li} analyzed the bound states of neutral fermions
interacting with homogeneous and linear electric fields without realizing,
in fact, that he was treating in the last case with the Dirac oscillator.
Indeed, the magnetic anomalous interaction has the form $-i\mu \beta \vec{%
\alpha}.\vec{\nabla}\phi (r)$, where $\mu $ is the anomalous magnetic moment in units
of the Bohr magneton and $\phi $ is the electric potential, \textit{i.e.}, the time
component of a vector potential \cite{tha}. Therefore, it is not hard to convince
oneself, without resort to a Schr\"{o}dinger-like equation and its nonrelativistic
limit, that the interaction $-im\omega \beta \vec{r}$ could come from a harmonic
oscillator electric potential.

The one-dimensional Dirac equation developed in this Letter could also be
obtained from the four-dimensional one with the mixture of spherically
symmetric scalar, vector and anomalous magnetic interactions. If we had
limited the fermion to move in the $x$-direction ($p_{y}=p_{z}=0$) the
four-dimensional Dirac equation would decompose into two equivalent
two-dimensional equations with 2-component spinors and 2$\times $2 matrices.
Then, there would result that the scalar and vector interactions would
preserve their Lorentz structures whereas the anomalous magnetic interaction
would turn out to be a pseudoscalar interaction. Therefore, the problem
presented in this Letter could be considered as that one of confinement of
neutral fermions by piecewise homogeneous (double-stepped) electric fields
in such a way to produce a bowl-shaped electric potential. The previous
argument, in addition to the former remark on the effective mass, provides a
way of looking at the those sorts of problems and makes the confinement of
neutral fermions in pseudoscalar potentials more plausible.

\bigskip

\smallskip

\noindent \textbf{Acknowledgments}

This work was supported in part through funds provided by CNPq and FAPESP.

\newpage

\end{document}